\documentclass[12pt]{iopart}
\usepackage{cite}

\usepackage{iopams}  
\usepackage{graphicx}
\usepackage{subfigure}
\usepackage[colorlinks,
    linkcolor=blue,
    anchorcolor=blue,
    citecolor=blue
]{hyperref}
\usepackage{makecell}
\usepackage{soul}
\begin{document}

\title{I-mode Plasma Confinement Improvement by Real-time Lithium Injection and its Classification on EAST Tokamak}
\renewcommand{\thefootnote}{\fnsymbol{footnote}}
\author{X.M. Zhong$^{1,2}$, X.L. Zou$^{3}$, A.D. Liu$^{\ast4}$, Y.T. Song$^{\ast5}$, G. Zhuang$^{4}$, H.Q. Liu$^{5}$, L.Q. Xu$^{5}$, E.Z. Li$^{5}$, B. Zhang$^{5}$, G.Z. Zuo$^{5}$, Z. Wang$^{5}$, C. Zhou$^{4}$, J. Zhang$^{4}$,  W.X. Shi$^{4}$, L.T. Gao$^{4}$, S.F. Wang$^{4}$, W. Gao$^{5}$, T.Q. Jia$ ^{5}$, Q. Zang$^{5}$, H.L. Zhao$^{5}$, M. Wang$^{5}$, H.D. Xu$^{5}$, X.J. Wang$^{5}$, X. Gao$^{1,2,5}$, X.D. Lin$^{1,2}$, J.G. Li$^{1,2,5}$, EAST I-mode Working Group$^{a}$, and EAST Team$^{b}$.} 
\address{$^{1}$College of Physics and Optoelectronic Engineering, Shenzhen University, Shenzhen 518060, China}
\address{$^{2}$Advanced Energy Research Center, Shenzhen University, Shenzhen 518060, China}
\address{$^3$ CEA, IRFM, F-13108 St Paul Les Durance, France}
\address{$^{4}$School of Nuclear Science and Technology, University of Science and Technology of China, Anhui Hefei 230026, China}
\address{$^{5}$Institute of Plasma Physics, Chinese Academy of Sciences, Anhui Hefei 230021, China}
\address{$\ast$ Authors to whom any correspondence should be addressed}
\address{$a$  See Liu et al 2024 (https://doi.org/10.1088/1741-4326/ad0acd) for the EAST I-mode Working Group}
\address{$b$  See Wan et al 2017 (https://doi.org/10.1088/1741-4326/aa7861) for the EAST Team}
\ead{lad@ustc.edu.cn, songyt@ipp.ac.cn}
\vspace{10pt}
\begin{indented}
\item[]\today
\end{indented}

\begin{abstract}
I-mode is a promising regime for future fusion reactors due to the high energy confinement and the moderate particle confinement.
However, the effect of lithium, which has been widely applied for particle recycling and impurity control, on I-mode plasma is still unclear.
Recently, experiments of real-time lithium powder injection on I-mode plasma have been carried out in EAST Tokamak.
It was found that the confinement performance of the I-mode can be improved by the lithium powder injection, which can strongly reduce electron turbulence (ET) and then trigger ion turbulence (IT).
Four different regimes of I-mode have been identified in EAST.
The Type I I-mode plasma is characterized by the weakly coherent mode (WCM) and the geodesic-acoustic mode (GAM).
The Type II I-mode is featured as the WCM and the edge temperature ring oscillation (ETRO).
The Type III I-mode corresponds to the plasma with the co-existence of ETRO, GAM, and WCM.
The Type IV I-mode denotes the plasma with only WCM but without ETRO and GAM.
It has been observed that WCM and ETRO are increased with lithium powder injection due to the reduction of ion and electron turbulence, and the enhancement of the pedestal electron temperature gradient.
EAST experiments demonstrate that lithium powder injection is an effective tool for real-time control and confinement improvement of I-mode plasma.

\item Key words: Confinement improvement, I-mode classification, Lithium injection, Turbulence.
\end{abstract}

\section{Introduction}

I-mode, which characterizes a temperature pedestal without a density pedestal, is a promising plasma regime for future fusion reactors due to the absence of edge-localized modes (ELMs)\cite{whyte2010mode}.
For fusion reactors, the lack of the edge particle transport barrier allows the I-mode to avoid the accumulation of plasma core impurities and facilitates helium ash removal. 
Recently, I-mode plasma has been realized on several Tokamaks, such as Alcator C-Mod\cite{greenwald1997h, hubbard2011edge}, ASDEX-U\cite{ryter1998h, happel2019stationarity, ryter2016mode}, DIII-D\cite{marinoni2015characterization}, EAST\cite{feng2019mode}, HL-2A\cite{liang2023identification}, and so on.
Generally, I-mode is usually obtained in the unfavorable configuration, i.e., the $B \times \nabla B$ ion drift pointing away from the active X-point.
I-mode is usually accompanied by the weakly coherent mode (WCM), which is localized in the pedestal region. 
In Alcator C-Mod, the geodesic-acoustic mode (GAM) may play a key role during the L-I-H transition due to nonlinear flow-turbulence coupling\cite{cziegler2017turbulence}.
A low frequency oscillation near the Last Closed Flux Surface (LCFS), which is referred to as the Low Frequency Edge Oscillation (LFEO), is observed in the Alcator C-Mod and ASDEX-U I-mode plasma\cite{mccarthy2022low, bielajew2022edge}.
LFEO is thought to be a type of GAM, and it is believed to be unnecessary for I-mode operation and may play a role in the regulation of particle transport\cite{mccarthy2022low}.
During the transition from I-mode to H-mode, a relaxation of both edge temperature and density profiles was observed in Alcator C-Mod and ASDEX-U, which is referred to as pedestal relaxation events (PREs)\cite{silvagni2022mode, hubbard2016multi, silvagni2020mode}.
A simulation\cite{manz2021gyrofluid} shows that a growing oscillation close to the separatrix provokes the PREs.

Recently, the stationary I-mode plasma has been identified in EAST, accompanied by a low-frequency coherent mode of $6–12 kHz$, called the edge temperature ring oscillation (ETRO)\cite{liu2020experimental}.
The alternating transition between the ion diamagnetic drift turbulence and the electron diamagnetic drift turbulence allows the I-mode to be maintained\cite{liu2020experimental}.
Furthermore, it was demonstrated that ETRO is not GAM, and the ETRO frequency is always two to three times lower than that of GAM\cite{liu2023characteristics}.
Moreover, the stationary I-mode with helium plasma was obtained in\cite{zhang2021mode}. 
In addition, pedestal burst instability (PBI), which is probably caused by density gradient, was observed during the I-H transition in EAST\cite{zhong2022characterization}.
The appearance of PBI and the prompt increase of density gradient before PBI allows for identifying the precursor for controlling I-H transition\cite{zhong2022characterization}.
Furthermore, the blob characteristics of the I-mode on EAST are similar to those of the L-mode but different from those of the H-mode, which might be due to the difference in collisionality in the scrape-off layer (SOL) region\cite{ping2023blob}.
In 2021, a 1000s Super I-mode plasma with double transport barriers combining electron-internal transport barrier (e-ITB) and edge electron heat transport barrier (I-mode) was discovered on EAST\cite{song2023realization}.
In Super I-mode, the use of real-time lithium powder injection suppresses impurities and reduces the edge recycling level.
However, so far, the effect of lithium on I-mode plasma is still unclear.

As a low-Z plasma-facing material, lithium has been widely applied for the conversion of properties in the first wall in several Tokamak devices, such as TFTR\cite{kugel1997development}, EAST\cite{zuo2010primary, lunsford2018injected, li2020development}, DIII-D\cite{bortolon2016high, bortolon2017mitigation}, ASDEX-U\cite{lang2016impact}, NSTX\cite{scotti2015lithium, kugel2012nstx, kugel2001overview}, and so on.
In EAST, lithium applications can be broadly classified into the following three categories: coating, injection, and flowing liquid lithium limiter (FLiLi)\cite{hu2023review}.
Lithium coating pretreatment effectively reduces the impurity radiation, the effective ion charge ($Z_{eff}$), the particle recycling level, and the H/(H$+$D) ratio, thereby improving density control, ICRF heating efficiency, and plasma confinement performance.
However, the thickness of the lithium film deposition is limited, therefore the service life is short.
In H-mode plasma, lithium powder injection enhances the edge coherent mode (ECM) intensity, which increases the edge particle and thermal transport and suppresses the ELM explosion\cite{hu2015new}.
And the application of FLiLi halves the heat flux on the limiter\cite{zuo2017mitigation}.
In addition, it is found that the edge electron temperature is increased with the application of lithium surfaces in LTX-$\beta$\cite{boyle2023extending}.
Currently, most stationary I-mode discharges in EAST are obtained under lithium wall conditions, but the influence of lithium on pedestal turbulence is still unknown.
In this work, the effect of lithium powder injection on I-mode plasma confinement performance and the pedestal turbulence will be investigated in detail.

The paper is organized in the following way.
Section 2 presents the experiment setup.
In section 3, the I-mode performance improvement with lithium powder injection is reported.
The effect of lithium powder injection on pedestal turbulence is shown in section 4.
Identification of four types of I-mode plasma in EAST is presented in section 5.
Finally, section 6 is the conclusion.

\section{Experiment setup}
\begin{figure}[htbp]
\centering
\includegraphics[width=3.5in]{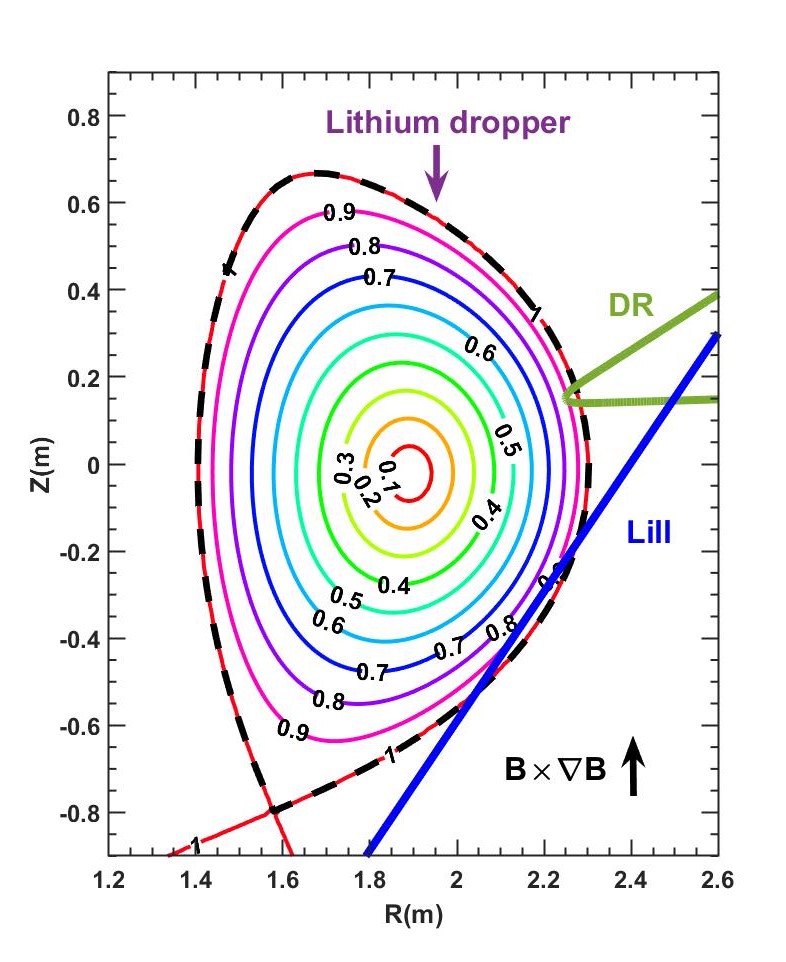}
\caption{Setup of lithium dropper in I-mode experiment.}
\label{DR_LiII_layout}
\end{figure}
Recently, the Lithium powder injection experiment for improving I-mode plasma confinement was carried out in a  lower single null (LSN) configuration on EAST, where the toroidal magnetic $B_T $ is $ 2.47T$, the plasma current $I_p $ is $450kA$, the maximum radius $R$ is $1.9m$, the minor radius $a$ is $ 0.45m$. 
The auxiliary heating methods are lower hybrid current drive (LHCD) and electron cyclotron resonance heating (ECRH).
Figure \ref{DR_LiII_layout} shows the equilibrium magnetic flux surfaces.
The normalized toroidal flux ($\rho$) is shown in the contour line.
Micron-sized lithium powder particles are injected into the plasma vacuum chamber by gravity through a dropper controlled by a dual piezoelectric crystal shaker\cite{canik2018active}. 
By adjusting the crystal voltage, the rate of lithium powder injection can be controlled.
The position of the lithium powder dropper is $R \sim 1.95m, Z \sim0.6m$, as shown in Figure \ref{DR_LiII_layout}.
Driven by gravity, the lithium powder enters into the SOL region at a speed of approximately $9 m/s$.
Lithium is rapidly liquefied, vaporized, and ionized to $Li^{+}$ by electron collision heating in the SOL region, emitting the green LiII line ($\lambda=548.3nm$).
The absolute intensity of the LiII line emission can be measured locally by the Filterscope system\cite{xu2016filterscope}, of which the chord viewing poloidally is tangent to the $\rho \sim 0.9$ magnetic surface, as shown in Figure \ref{DR_LiII_layout} (blue line).
By launching the probing beam at an oblique angle and receiving the backscattered signal by the fluctuation around the cut-off layer, the turbulence can be measured with high spatial and temporal resolution by the Doppler Reflectometer (DR)\cite{zhou2013microwave}.
The DR phase derivative perturbation ($\frac{d\widetilde{\phi }}{dt}$) includes $E \times B$ velocity perturbation ($ \widetilde{V}_{E\times B}$), turbulence phase velocity perturbation ($\widetilde{V}_{phase}$), and the phase modulation by cut-off oscillation ($ \frac{d\widetilde{\phi }_0}{dt}$), {i.e., $\frac{d\widetilde{\phi }}{dt}= k_\perp (  \widetilde{V}_{E\times B} + \widetilde{V}_{phase} )  +  \frac{d\widetilde{\phi }_0}{dt} $}, where $k_\perp$ corresponds to the turbulence perpendicular wavenumber\cite{zou1999poloidal, hillesheim2010new, fanack1996ordinary}.
Therefore, in DR $d\widetilde{\phi } / dt$ spectra, the WCM can be measured by the phase modulation, ETRO can be measured by the turbulence phase velocity perturbation, and GAM can be measured by the $E \times B$ velocity perturbation\cite{zhong2022characterization}. 
In the experiment, the cut-off layer for DR ($74GHz$) is about $\rho \sim 0.9$ by the ray tracing calculation, as shown in Figure \ref{DR_LiII_layout} (green line).
The electron density is measured by hydrogen cyanide (HCN) laser interferometer\cite{shen2013improved}, polarimeter-interferometer (POINT)\cite{liu2014faraday}, and reflectometer\cite{zhang2023determination}. 
In addition, the electron temperature is measured by Thomson scattering (TS)\cite{qing2010development}, electron cyclotron emission (ECE)\cite{liu2019overview}, and correlation electron cyclotron emission (CECE)\cite{zhao2019multi}.

\section{I-mode performance improvement with lithium injection}

\begin{figure}[htbp]
\centering
\includegraphics[width=5in]{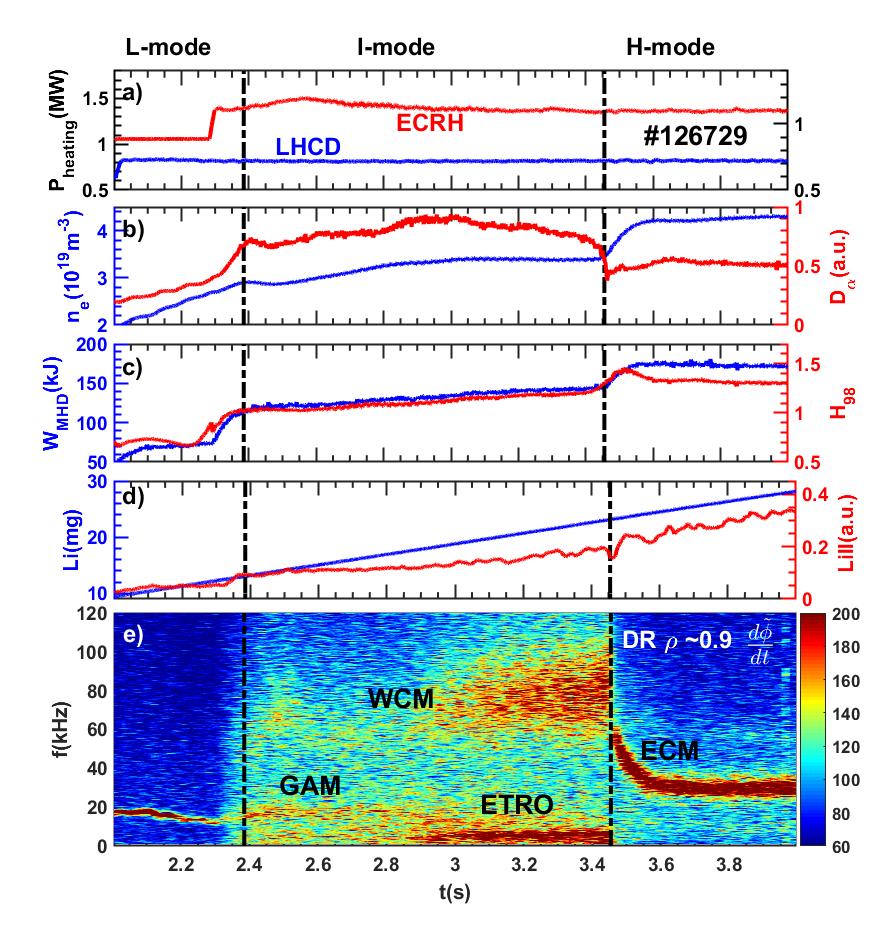}
\caption{Lithium powder injection in EAST I-mode plasma (shot $\#126729$). a) LHCD power (blue line) and ECRH power (red line). b) Chord-averaged density (blue line) and $D_\alpha$ signal. c) plasma stored energy $W_{MHD}$ and $H_{98} factor$. d) Mass of lithium powder injected (blue line) and LiII radiation (red line). e) Time-frequency spectrogram of the DR phase derivative signal.}
\label{result126729}
\end{figure}

The experimental results of Li powder injection (shot$\#126729$) are displayed in Figure \ref{result126729}, with $I_p \sim 450kA$, $P_{LHCD} \sim 0.8MW$, $P_{ECRH} \sim 1.4MW$.
As the auxiliary heating power increases, the L-I transition at $t \sim 2.385s$ can be identified by the appearance of the WCM in the {$d\widetilde{\phi } / dt$} spectrogram, as well as the increase of the plasma stored energy $W_{MHD}$ and the $H_{98}$ factor. 
Throughout the discharge, the Li powder is continuously injected into the plasma at a rate of $9.4 mg/s$, which is consistent with the increase in LiII radiation intensity, as shown in Figure \ref{result126729} d.
With the lithium powder injection, the intensity of WCM is increased, while ETRO appears and GAM disappears.
The I-mode plasma transits to the H-mode plasma at $t \sim 3.455s$, which can be identified by the sudden decrease of the $D_{\alpha}$ intensity, as well as the appearance of the edge coherent mode (ECM) in the $d\widetilde{\phi } / dt$ spectrogram.
Meanwhile, the chord-averaged density, the plasma stored energy, and the $H_{98}$ factor increase significantly during the I-H transition.

\begin{figure}[htbp]
\centering
\includegraphics[width=5.5in]{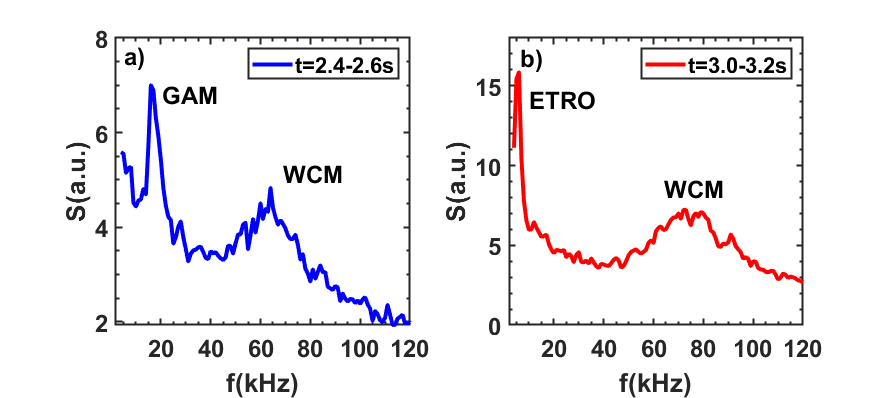}
\caption{$d\widetilde{\phi } / dt$ power spectrum of DR in the pedestal region during I-mode plasma.}
\label{dphidt126729}
\end{figure}

Figure \ref{dphidt126729} shows the $d\widetilde{\phi } / dt$ power spectrum of DR in the pedestal region during the I-mode plasma.
From Figure \ref{result126729} e, the I-mode plasma can be divided into 2 distinct regimes: one with only GAM and WCM as shown in Figure \ref{dphidt126729} a, and another one with only ETRO and WCM as shown in Figure \ref{dphidt126729} b.
With the injection of lithium powder and the increase of plasma energy, the center frequency of the WCM is increased significantly from $65 kHz$ to $75 kHz$, as does the intensity of the WCM.

\begin{figure}[htbp]
\centering
\includegraphics[width=5.5in]{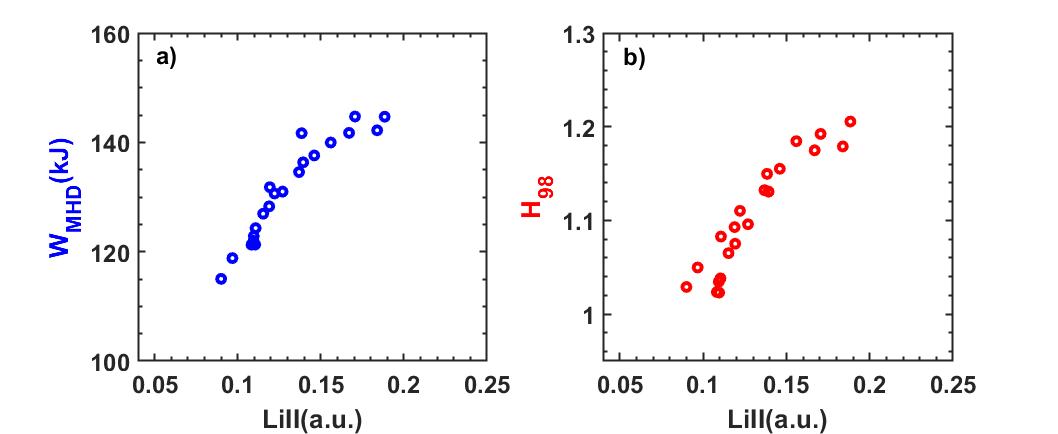}
\caption{LiII radiation, versus the plasma stored energy a) and $H_{98}$ factor, respectively.}
\label{li126729}
\end{figure}

As shown in Figure \ref{result126729} d, the LiII radiation is gradually increased by injecting the Li powder during the I-mode regime.
Figure \ref{li126729} shows that both plasma stored energy and $H_{98}$ factor are increased with LiII radiation, indicating that the confinement performance of the I-mode can be strongly improved by the lithium powder injection.

\section{Effect of lithium injection on pedestal turbulence}

\begin{figure}[htbp]
\centering
\includegraphics[width=5in]{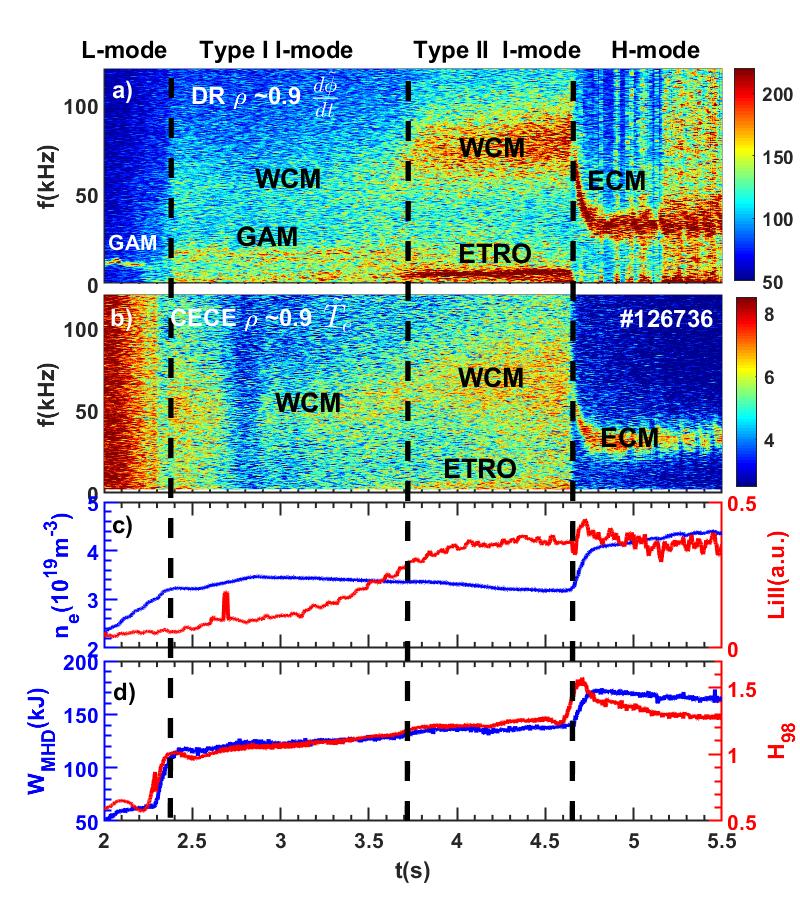}
\caption{Transition from Type I I-mode to Type II I-mode with lithium powder injection in shot $\#126736$. a) Time-frequency spectrogram of $d\widetilde{\phi } / dt$ in the pedestal region. b) Spectrogram of temperature perturbation by CECE. c) Chord-averaged density (blue line) and LiII radiation (red line). d) plasma stored energy $W_{MHD}$ and $H_{98} factor$.}
\label{result126736}
\end{figure}

For simplicity, we refer to the I-mode with only GAM and WCM as the Type I I-mode.
And the I-mode with only ETRO and WCM is called the Type II I-mode.
Figure \ref{result126736} a is the time-frequency spectrogram of $d\widetilde{\phi } / dt$ measured by DR, Figure \ref{result126736} b is the spectrogram at the temperature perturbation measured by CECE, Figure \ref{result126736} c shows the chord-averaged density and the LiII radiation, Figure \ref{result126736} d shows the plasma stored energy $W_{MHD}$ and the $H_{98}$ factor.
From the DR spectrogram (Figure \ref{result126736} a), the whole discharge of shot $\#126736$ can be divided into four phases. 
Interval 1 is the L-mode, where GAM with the frequency of $12 kHz$  is clearly observed.
Interval 2 corresponds to the Type I I-mode, where a GAM with an increased frequency of $17 kHz$ can be observed, which can be explained by the increase of the temperature.
In interval 3, the Type II I-mode emerges with the appearance of the ETRO, accompanied by an increase in the amplitude and frequency of the WCM, and the disappearance of the GAM.
Interval 4 is the H-mode with the emergence of the ECM.
From the CECE spectrogram, ETRO, WCM, and ECM can be clearly observed, while GAM is not observed.
This is due to the fact that GAM is a $E \times B$ velocity perturbation, not a temperature perturbation.
With the lithium powder injection, LiII radiation is increased gradually and then reaches a saturated situation due to the lithium transport.




\begin{figure}[htbp]
\centering
\subfigure{
\includegraphics[width=2.5in]{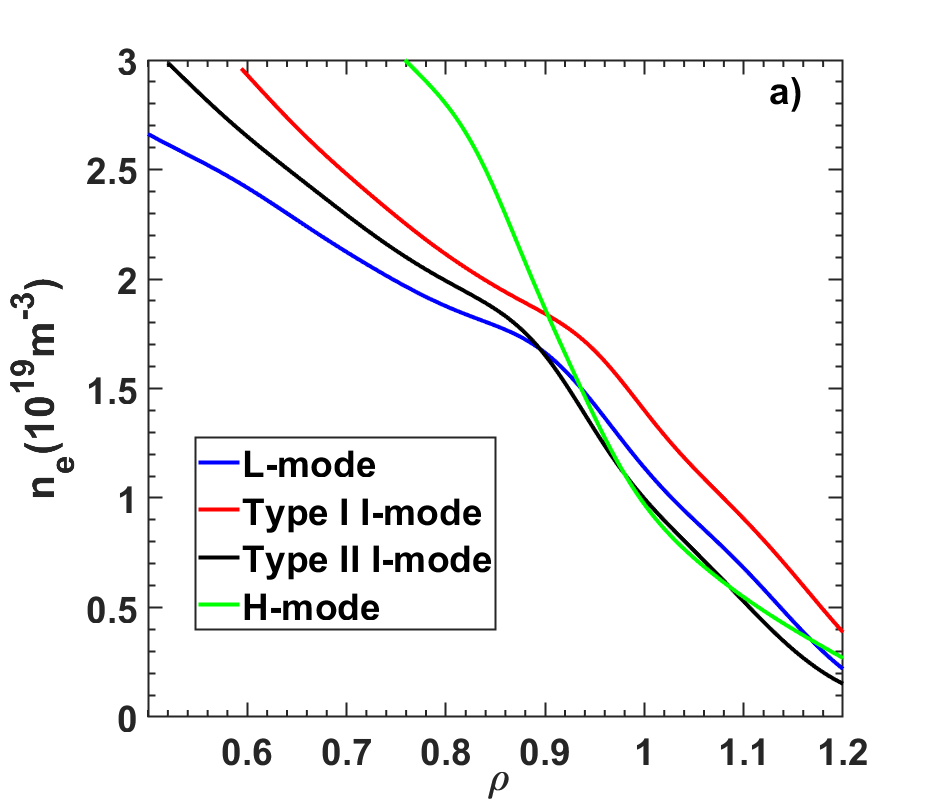}
}
\subfigure{
\includegraphics[width=2.5in]{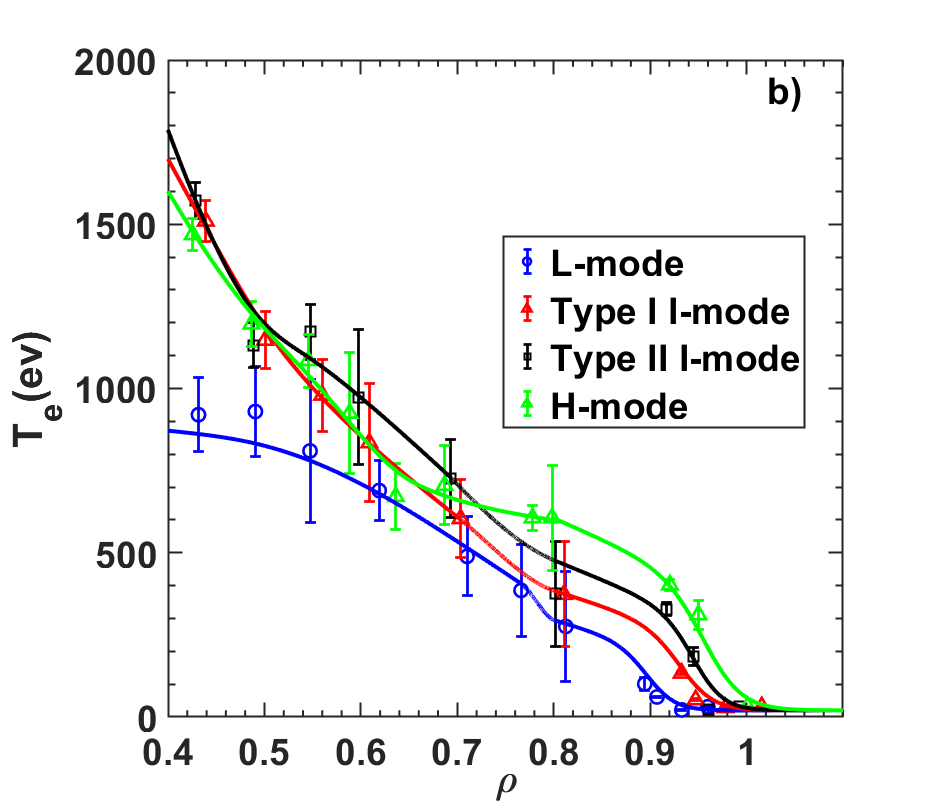}
}

\caption{Profiles of electron density a) measured by reflectometry and electron temperature b) measured by TS, and ECE during the L-mode, the Type I I-mode, the Type II I-mode, and the H-mode, respectively.}
\label{profile126736}
\end{figure}

The profiles of electron density measured by reflectometry and electron temperature measured by TS, and ECE during the L-mode, the Type I I-mode, the Type II I-mode, and the H-mode are shown in Figure \ref{profile126736} a and Figure \ref{profile126736} b, respectively.
Note that a distinct density pedestal is observed in the H-mode plasma, located at the $\rho \sim 0.9$.
In contrast, the density profiles of the Type I I-mode and Type II I-mode do not exhibit the density pedestal, like as the L-mode.
Meanwhile, a significant electron temperature pedestal is observed in both Type I and II I-mode, when compared to the L-mode.
An interesting point is that the Type II I-mode has a higher temperature pedestal than that of the Type I I-mode and a lower temperature pedestal than that of the H-mode.
The enhanced temperature gradient at the pedestal region suggests that turbulence has been reduced due to the real-time lithium powder injection, which will be investigated in detail later.

\begin{figure}[htbp]
\centering
\includegraphics[width=6.5in]{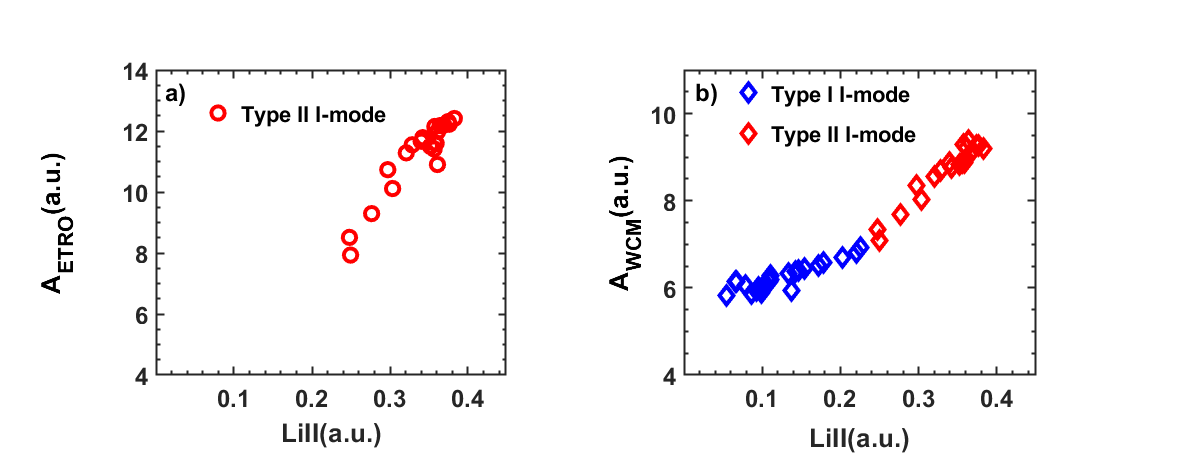}
\caption{LiII radiation, versus ETRO intensity a) and WCM intensity b).}
\label{li126736}
\end{figure}


\begin{figure}[htbp]
\centering
\subfigure{
\includegraphics[width=3.5in]{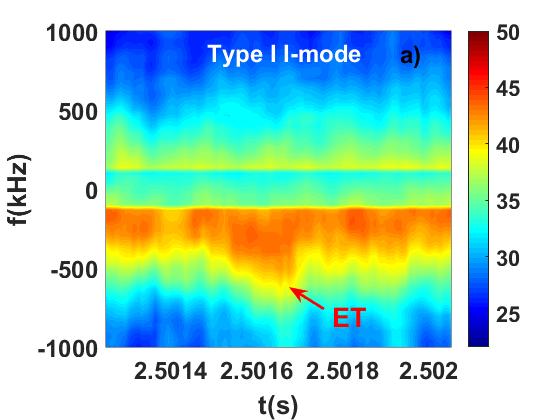}
}
\subfigure{
\includegraphics[width=3.5in]{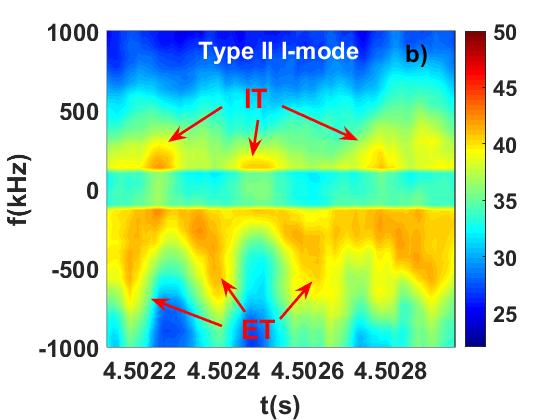}
}
\subfigure{
\includegraphics[width=3.2in]{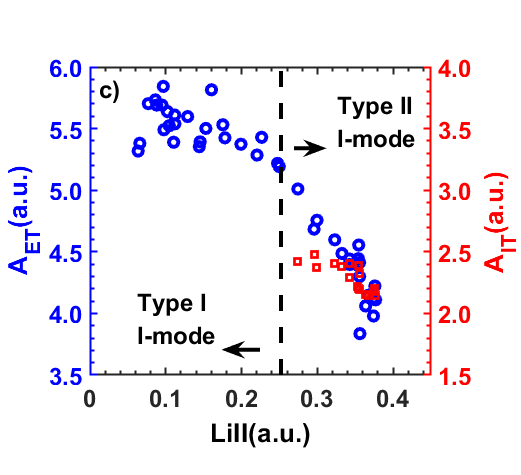}
}
\caption{Turbulence spectrogram during Type I I-mode a) and Type II I-mode b). c) LiII radiation, versus the electron turbulence (ET) intensity and the ion turbulence (IT) intensity.}
\label{ETIT126736}
\end{figure}

Figure \ref{li126736} a and Figure \ref{li126736} b show that both WCM and ETRO intensities are increased with LiII radiation, where the integrating frequency range is from $5-7 kHz$ for ETRO, and $40 -100kHz$ for WCM.
This phenomenon could be explained by the pedestal turbulence change.
The turbulence spectrogram measured by DR during the Type I and II I-mode are shown in Figure \ref{ETIT126736} a and Figure \ref{ETIT126736} b, respectively.
In the turbulence spectrogram, we refer to the electron turbulence as simply ET, and the ion turbulence as simply IT.
ET and IT express electron/ion diamagnetic drift direction turbulence, respectively.
During the Type I I-mode plasma, only ET is observed in the turbulence spectrogram.
In contrast, both ET and IT are present in the Type II I-mode.
In addition, there appears to be an alternating transition between ET and IT, which is consistent with the previously reported physical mechanisms of the ETRO.
Figure \ref{ETIT126736} c shows the reduction in ET intensity as LiII increases.
The I-mode plasma transits from Type I to Type II when IT is triggered.
It is noted that both ET and IT intensities decrease with LiII radiation, indicating that the pedestal turbulence can be strongly reduced by the lithium powder injection, which can explain the reason for the confinement improvement with lithium powder injection.


\section{Classificaiton of I-mode in EAST}

\begin{figure}[htbp]
\centering
\includegraphics[width=4.2in]{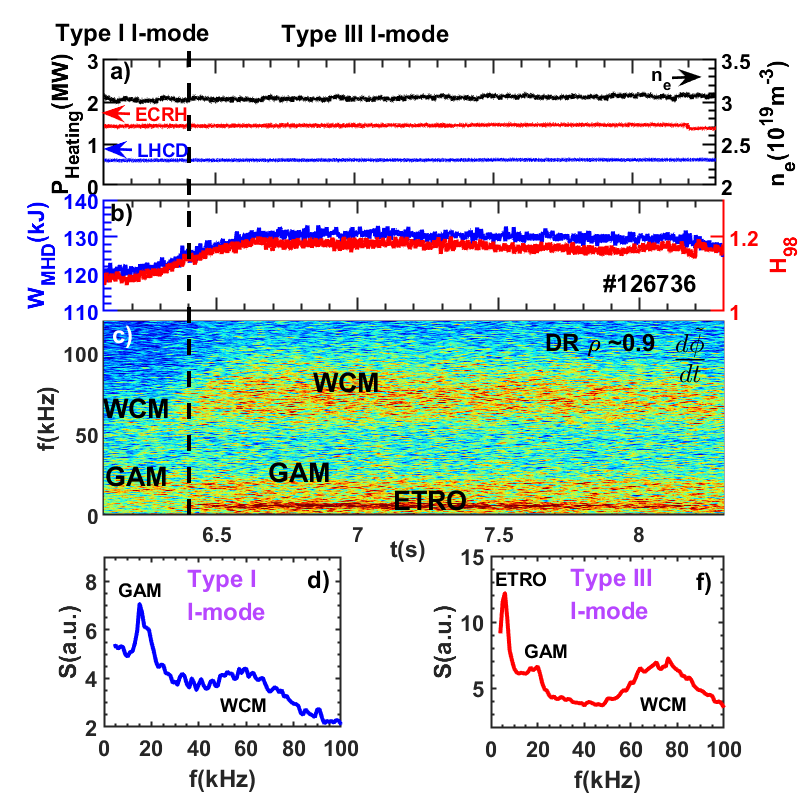}
\caption{Transition from Type I I-mode to Type III I-mode. a) Heating power of  LHCD (blue line) and ECRH (red line) and the chord-average density (black line). b)  Plasma stored energy $W_{MHD}$ (blue line) and $H_{98}$ factor. c) $d\widetilde{\phi } / dt$ spectrogram of DR in the pedestal region. d) $d\widetilde{\phi } / dt$ power spectrum during Type I I-mode. e) d) $d\widetilde{\phi } / dt$ power spectrum during Type III I-mode.}
\label{Type I_III_transition}
\end{figure}

\begin{figure}[htbp]
\centering
\includegraphics[width=4.2in]{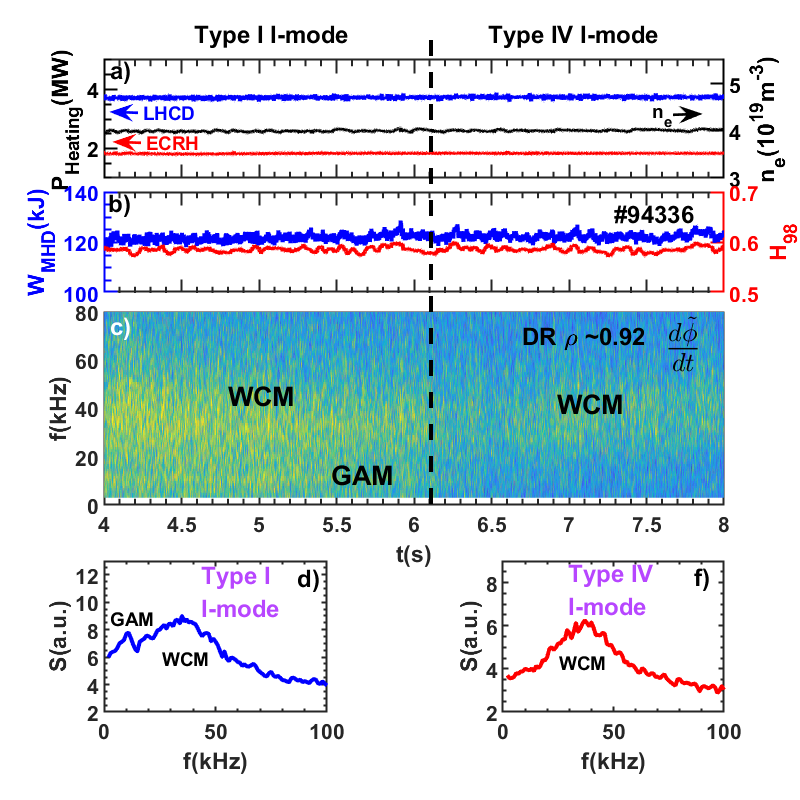}
\caption{Transition from Type I I-mode to Type IV I-mode. a) Heating power of  LHCD (blue line) and ECRH (red line) and the chord-average density (black line). b)  Plasma stored energy $W_{MHD}$ (blue line) and $H_{98}$ factor. c) $d\widetilde{\phi } / dt$ spectrogram of DR in the pedestal region. d) $d\widetilde{\phi } / dt$ power spectrum during Type I I-mode. e) d) $d\widetilde{\phi } / dt$ power spectrum during Type IV I-mode.}
\label{Type I_IV_transition}
\end{figure}

In addition to the Type I and II I-mode mentioned above, there are other types of I-mode in EAST.
For simplicity, we call the plasma with ETRO, GAM, and WCM the Type III I-mode.
The plasma with WCM only is referred to as the Type IV I-mode.
Figure \ref{Type I_III_transition} shows the transition from Type I I-mode to Type III I-mode in shot $\#126736$.
During the Type III I-mode plasma, it can be clearly seen that there exists an ETRO with a frequency of $5kHz$, a GAM with a frequency of $17kHz$ and a WCM with a frequency range of about $50-100kHz$.
It is observed that the $H_{98}$ factor is increased by about $10\%$ when ETRO is present.
Figure \ref{Type I_IV_transition} shows the transition from Type I I-mode to Type IV I-mode in shot $\#94336$ with the helium plasma.
Only WCM is observed during the Type IV I-mode plasma, without GAM and ETRO.
It should be noted that there is no significant change in the chord-averaged density and the plasma stored energy when the I-mode plasma transits from Type I to Type IV.

																													Classification of I-mode plasma in EAST is listed in Table \ref{I-mode}.
The Type I I-mode plasma is characterized by the WCM and the GAM.
The Type II I-mode is featured as the WCM and the ETRO.
The Type III I-mode corresponds to the plasma with the co-existence of ETRO, GAM, and WCM.
The Type IV I-mode is defined as the plasma with only WCM but without ETRO and GAM.
In ASDEX-U and Alcator C-Mod, the LFEO is observed in $40-60\%$ I-mode discharges\cite{mccarthy2022low}.
Since LFEO is considered as a GAM\cite{mccarthy2022low}, these plasmas belong to the Type I I-mode.
ETRO is essential for EAST stationary I-mode discharges\cite{liu2020experimental, liu2023characteristics}, so the EAST stationary I-mode is more likely a Type II I-mode.
The I-mode plasma observed on HL-2A should be the Type IV I-mode, as only WCM is present while GAM and ETRO are absent\cite{liang2023identification}.

 \begin{table}[htbp]
 \caption{{Classification of I-mode in EAST}.}
\label{I-mode}
 \begin{center}
 \begin{tabular}{ c c c c c}
\hline
&\makecell [c]{Type I \\I-mode}&\makecell [c]{Type II \\I-mode}&\makecell [c]{Type III \\I-mode}&\makecell [c]{Type IV \\I-mode}\\

WCM &$\surd$  &$\surd $  &$\surd $  &$\surd $\\

ETRO &  &$\surd $  &$\surd $  &\\

GAM &$\surd $  &   &$\surd $  &\\
\hline
 \end{tabular}
 \end{center}
 \end{table}

\section{Conclusions}

Recently, experiments of real-time lithium powder injection on I-mode plasma have been carried out in EAST Tokamak.
It was found that the confinement performance of the I-mode can be improved by the real-time lithium powder injection.
In addition, a transition from Type I I-mode to Type II I-mode is observed with lithium powder injection, where Type I I-mode is featured as WCM and GAM, and Type II I-mode is defined as the plasma with WCM and ETRO.
The Type I I-mode has a higher temperature pedestal than that of the L-mode but a lower temperature pedestal than that of the Type II I-mode.
During the Type I I-mode, only electron turbulence is present, while both electron turbulence and ion turbulence are observed during the Type II I-mode.
The I-mode plasma transits from Type I to Type II when ion turbulence is triggered.
The pedestal turbulence (both electron turbulence and ion turbulence) can be strongly reduced by the lithium powder injection.
Both WCM and ETRO intensities are increased with lithium powder injection due to the reduction of turbulence and the enhancement of the pedestal electron temperature gradient.
Additionally, another two types of I-mode have been identified in EAST.
The Type III I-mode corresponds to the plasma with the co-existence of ETRO, GAM, and WCM.
The Type IV I-mode is defined as the plasma with only WCM but without ETRO and GAM.
EAST experiments demonstrate that lithium powder injection is an effective tool for real-time control and confinement improvement of I-mode plasma.
However, the physical mechanism of the turbulence reduction by lithium powder injection and the formation mechanism of Type IV I-mode are still unclear and will be further investigated in future work.

\section*{Acknowledgments:}
This work was supported by the National Key R$\&$D Program of China under Grant Nos. 2022YFE03070000, 2022YFE03070004, the Natural Science Foundation of China under Grant Nos. 12075155, U1967206, 11975231, and 11922513, the National MCF Energy R$\&$D Program under Grant Nos. 2017YFE0301204 and 2018YFE0311200, the Users with Excellence Program of Hefei Science Center CAS under Grant No. 2020HSC-UE009 and Fundamental Research Funds for the Central Universities.  
We also acknowledge the EAST team for supporting experiments.

\section*{Reference:}

\bibliographystyle{unsrt}
\bibliography{ref}
\end{document}